\documentclass[conference]{IEEEtran}
\IEEEoverridecommandlockouts
\usepackage[letterpaper, left=0.62in, right=0.62in, bottom=1in, top=0.75in]{geometry}
\usepackage{amssymb,amsmath}
\usepackage{graphicx}
\usepackage{cite}
\usepackage{txfonts}
\usepackage{mathrsfs}
\usepackage{fontenc}
\usepackage[binary-units]{siunitx}
\DeclareSIUnit{\bps}{bps}

\usepackage[caption=false,font=footnotesize]{subfig}
\usepackage{tabularx}
\usepackage{multirow}
\usepackage{diagbox}
\usepackage{algorithm2e}
\usepackage{bbm}
\usepackage{flushend}
\usepackage{xcolor}

\begin{document}

\title{A Simple Multiple-Access Design for Reconfigurable Intelligent Surface-Aided Systems}

\author{\IEEEauthorblockN{Wei Jiang\IEEEauthorrefmark{1} and Hans D. Schotten\IEEEauthorrefmark{2}}
\IEEEauthorblockA{\IEEEauthorrefmark{1}German Research Center for Artificial Intelligence (DFKI)\\Trippstadter Street 122,  Kaiserslautern, 67663 Germany\\
  }
\IEEEauthorblockA{\IEEEauthorrefmark{2}Rheinland-Pf\"alzische Technische Universit\"at Kaiserslautern-Landau\\Building 11, Paul-Ehrlich Street, Kaiserslautern, 67663 Germany\\
 }
\thanks{$^{\divideontimes}$This work was supported in part by the European Commission H2020 Framework Programme through \textit{AI@EDGE} (Grant no. \emph{101015922}) and in part by the German Federal Ministry of Education and Research (BMBF) through \emph{Open6G-Hub} (Grant no.  \emph{16KISK003K}) and \emph{AI-NET PROTECT} (Grant no.  \emph{16KIS1283})}.
}
\maketitle

\begin{abstract}
This paper focuses on the design of transmission methods and reflection optimization for a wireless system assisted by a single or multiple reconfigurable intelligent surfaces (RISs).  The existing techniques are either too complex to implement in practical systems or too inefficient to achieve high performance. To overcome the shortcomings of the existing schemes, we propose a simple but  efficient approach based on \textit{opportunistic reflection} and \textit{non-orthogonal transmission}. The key idea is opportunistically selecting the best user that can reap the maximal gain from the optimally reflected signals via RIS. That is to say, only the channel state information of the best user is used for RIS reflection optimization, which can in turn lower complexity substantially. In addition, the second user is selected to superpose its signal on that of the primary user, where the benefits of non-orthogonal transmission, i.e., high system capacity and improved user fairness, are obtained.  Additionally, a simplified variant exploiting random phase shifts is proposed to avoid the high overhead of RIS channel estimation.
\end{abstract}

\IEEEpeerreviewmaketitle

\section{Introduction}
Recently, reconfigurable intelligent surface (RIS), a.k.a intelligent reflecting surface (IRS), has drawn much attention from both academia and industry. It is considered a high-potential technology for the upcoming Six-Generation (6G) mobile systems \cite{Ref_jiang2021road}.  Particularly, RIS is a planar meta-surface consisting of a large number of reflection elements.  A smart controller dynamically adjusts the phase shift of each element, thereby collaboratively achieving a programmable radio propagation environment for signal amplification or interference suppression.  Traditional technologies, such as \textit{dense and heterogeneous networking}, \textit{massive MIMO}, and \textit{relaying} can improve coverage and capacity effectively but incur high costs, much power consumption, and severe signal interference. In contrast, RIS not only reflects signals in a full-duplex and noiseless way but also provides a power-efficient and cost-efficient solution that supports green and sustainable performance growth in the 6G system since each RIS element is passive, lightweight, and cheap \cite{Ref_renzo2020smart}.

Previous studies have investigated different aspects for RIS-aided wireless communications, e.g., reflection optimization \cite{Ref_wu2019intelligent}, RIS channel estimation \cite{Ref_wang2020channel}, hardware constraints \cite{Ref_zhi2021uplink}, and the interplay with other technologies, such as MIMO \cite{Ref_hu2018beyond}, OFDM \cite{Ref_zheng2020intelligent}, and Terahertz communications \cite{Ref_jiang2022dualbeam}.
Most prior works focus on point-to-point communications, consisting of a base station (BS), a user, and a RIS, for the sake of analysis.  However, any practical wireless system needs to serve lots of users, imposing the need for multiple-access design.
In \cite{Ref_jiang2023capacity, Ref_zheng2020intelligent_COML, Ref_jiang2023orthogonal}, the researchers studied non-orthogonal multiple access (NOMA) in a single-RIS-aided system, in comparison with orthogonal multiple access (OMA). In addition, Ding and Poor proposed a simple NOMA design for RIS \cite{Ref_ding2020simple}. The authors of \cite{Ref_nadeem2021opportunistic} proposed opportunistic beamforming (OppBF) to avoid channel estimation in multi-user RIS. In \cite{Ref_jiang2023userscheduling}, user scheduling and passive beamforming for FDMA/OFDMA in a RIS system are discussed. The work in \cite{Ref_jiang2022multiuser} studied the effect of discrete phase shifts on multi-user RIS communications.
The impact of multi-user scheduling on the performance of RIS is reported in \cite{Ref_gan2021user}.
The authors of this paper have also analyzed and compared different RIS multiple-access schemes for multi-RIS-aided vehicular networks \cite{Ref_jiang2022intelligent}.

On the other hand, it is highly probable that a  system deploys multiple distributed RISs, and therefore restricting to a single RIS is not sufficiently practical, as studied in  \cite{Ref_do2021multiRIS}. From a practical point of view, this paper considers a more generalized scenario, where a BS serves multiple users with the aid of either one or multiple RIS surfaces. While the aforementioned works have deepened our understanding, how to efficiently optimize RIS reflection in a multi-user multi-RIS-aided system is still an open issue. Through an exhaustive search, three applicable methods have been found. The simplest way is to assign users to orthogonal time-frequency resource pieces by time-division multiple access (TDMA) \cite{Ref_zheng2020intelligent_COML} or (orthogonal) frequency-division multiple access (FDMA/OFDMA). However, FDMA/OFDMA performs worse due to the lack of frequency-dependent phase shifts per RIS element \cite{Ref_wu2021intelligent}. But OMA including both TDMA and FDMA is inefficient where each user transmits over a small portion of available resources. Globally optimized reflection is achieved by joint RIS optimization (JRO) using semidefinite relaxation (SDR), which has been employed already for single-user single-RIS systems \cite{Ref_wu2019intelligent}. However, SDR causes a prohibitive computational complexity.  Selecting a single user opportunistically to transmit at each time substantially simplifies the RIS optimization. Based on the availability of channel state information (CSI), two variants, called opportunistic user selection (OUS)  \cite{Ref_gan2021user, Ref_jiang2023userSelection} and opportunistic beamforming (OppBF) \cite{Ref_nadeem2021opportunistic, Ref_dimitropoulou2022opportunistic}, have been designed. Nevertheless, these two schemes also suffer from substantial performance degradation in comparison with JRO. In \cite{Ref_jiang2023opportunistic}, opportunistic reflection with all-user transmission has been proposed. Although it achieves a good performance-complexity trade-off, it is vulnerable to hardware and channel impairments, such as channel aging and phase noise \cite{Ref_jiang2023performance}. Besides, all-user non-orthogonal transmission in \cite{Ref_jiang2023opportunistic} is hard to implement since the capability of the successive interference cancellation (SIC) at the receiver is limited by the number of superimposed users.

Responding to this, this paper aims to design a simple but efficient signal transmission and optimization design for multi-user multi-RIS systems. We combine \textit{opportunistic RIS reflection} with \textit{non-orthogonal transmission} and thereby design a novel scheme coined Opportunistic Reflection-based Non-Orthogonal Multiple Access (OR-NOMA) for RIS-aided wireless systems.  It selects an opportunistic user that can reap the maximal gain from the optimally reflected signals via RIS. That is to say, only the CSI of the selected user is needed for RIS reflection optimization, which can in turn substantially lower the complexity of RIS channel estimation and reflection optimization. In addition, the second user is (randomly or deliberately) selected to form a pair of active users to non-orthogonally transmit their signals simultaneously over the same time-frequency resource unit.  Thus, we get the non-orthogonal transmission gain, i.e., higher system capacity and improved user fairness. As a result, it substantially lowers the complexity of JRO by selecting an opportunistic user as the anchor for RIS optimization. Unlike OUS and OppBF, where only a selected user transmits its signal at a certain time, a pair of users are allowed to transmit simultaneously. Inspired by \cite{Ref_nadeem2021opportunistic, Ref_dimitropoulou2022opportunistic}, a simplified version of OR-NOMA exploiting random phase rotations is proposed to avoid the high overhead of RIS channel estimation \cite{Ref_liu2020matrix, Ref_zheng2021efficient, Ref_you2021reconfigurable}. Numerical results are applied to justify the proposed schemes.

\section{System Model}

\begin{figure}[!t]
    \centering
    \includegraphics[width=0.475\textwidth]{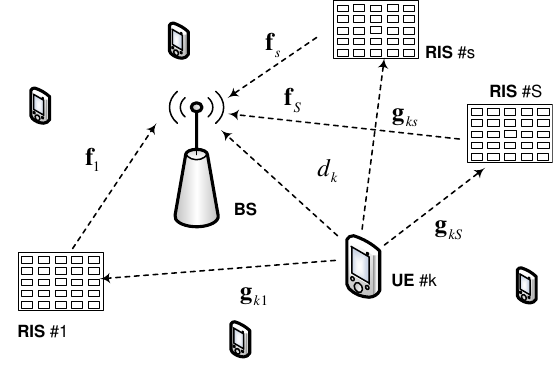}
    \caption{Schematic diagram of the uplink in a multi-RIS-aided multi-user communications system over \textit{i.n.i.d.} fading channels.  }
    \label{fig:SystemModel}
\end{figure}

This paper focuses on the uplink transmission of a wireless system, where $K\geqslant 1$ users access to a BS with the aid of $S\geqslant 1$ surfaces\footnote{The algorithm design and performance analysis are conducted under a generalized and more challenging setup of multiple RISs $S\geqslant1$. Simply setting $S=1$, it falls back to the single-RIS scenario, as assumed by most literature.}, as shown in \figurename \ref{fig:SystemModel}. Each RIS is equipped with a smart controller that adaptively adjusts the phase shift of each reflecting element based on the knowledge of instantaneous CSI.  The $s^{th}$ surface,  $s\in \{1,2,\ldots,S\}$, is comprised of $N_s$ elements, where a typical element $n_s\in \{1,2,\ldots,N_s\}$ is modeled as $\epsilon_{sn_s}=e^{j\theta_{sn_s}}$ with a phase shift $\theta_{sn_s}\in [0,2\pi)$.

In contrast to traditional wireless systems that rely on \textit{independent and identically distributed (i.i.d.)} Rayleigh or Rician channels, RIS-assisted communications encounter a diverse fading environment. To address this, a more versatile and applicable setup involving independently non-identically  distributed (\textit{i.n.i.d.}) conditions is required. Under this framework, each channel can exhibit unique fading characteristics. While mobile users typically experience multiple scattering effects, the connection between the base station and the RIS remains relatively stable, characterized by a line-of-sight (LOS) path due to fixed deployment locations. As a result, a more encompassing and practical \textit{i.n.i.d.} Nakagami-$m$ fading model is essential for correctly depicting RIS-assisted systems \cite{Ref_do2021multiRIS}.

Write $ X \sim \mathrm{Nakagami}(m, \Omega)$ to represent a random variable following Nakagami-$m$ distribution. Its probability
density function (PDF) and cumulative distribution function (CDF) are expressed as
\begin{equation}
    f_X(x)=\frac{2{m}^{m}}{\Gamma(m){\Omega}^{m}} x^{2m -1} e^{-\frac{m}{\Omega} x^2},
\end{equation}
and
\begin{equation}
    F_X(x)=\frac{\gamma \left(m, \frac{m}{\Omega} x^2 \right)}{ \Gamma(m)},
\end{equation}
respectively,
where $m$ indicates the severity of small-scale fading, $\Omega$ equals large-scale fading.
Rayleigh fading is a special case of Nakagami-\textit{m} fading if $m=1$.
Under the \textit{i.i.d.} assumption, the values of $m$ and $\Omega$ for different channels are identical. It is not suitable for RIS-aided communications where the fading severity and distance-dependent path losses are distinct.

Let $d_k\in \mathbb{C}$ denote the channel coefficient between user $k$ and the BS. We have $d_k=|d_k|e^{j\phi^d_{k}}$, where $\phi^d_{k}$ denotes phase and $|d_k|$ is magnitude, following Nakagami-\textit{m} fading, i.e., $|d_k| \sim \mathrm{Nakagami}\left(m^d_{k}, \Omega^d_{k} \right)$.
The channel between the $s^{th}$ RIS and the $k^{th}$ user is modeled as
\begin{equation}
    \mathbf{g}_{ks}=\Bigl[g_{k,s1},g_{k,s2},\ldots,g_{k,sN_s}\Bigr]^T,
\end{equation}
where $g_{k,sn_s}=|g_{k,sn_s}|e^{j\phi^g_{k,sn_s}}$ is the channel coefficient between the $n_s^{th}$ element of surface $s$ and user $k$. All elements over a RIS has the same propagation path to a certain user or the BS, so they experience similar, if not identical, fading statistics. Hence, we can apply $m^g_{sk}$ and $\Omega^g_{sk}$, $\forall s,k$ for the channels from RIS to users, i.e. $|g_{k,sn_s}| \sim \mathrm{Nakagami}\left(m^g_{sk}, \Omega^g_{sk}\right)$, $\forall n_s$.
Similarly, write
\begin{equation}
    \textbf{f}_s =[f_{s1},f_{s2},\ldots,f_{sN_s}]^T
\end{equation}
to denote the channel vector from the $s^{th}$ surface to the BS, where $|f_{sn_s}| \sim \mathrm{Nakagami}\left(m^f_s, \Omega^f_s\right)$, $\forall n_s$.

\section{Opportunistic Reflection-Based NOMA}
The straightforward approach involves partitioning the signaling dimension over the time axis into $K$ mutually orthogonal slots. Through round-robin scheduling, every user takes turns sequentially to access their designated slot. During the $k^{th}$ slot, a typical user labeled as $k$ transmits its symbol $s_k$, while the rest of the users remain silent. According to \cite{Ref_zhang2018space},  a RIS element employing positive-intrinsic-negative (PIN) diodes achieves a maximum switching frequency of \SI{5}{\mega\hertz}. This rate far surpasses the usual time slot adjustments, which occur at a scale of milliseconds (\si{\milli\second}). Consequently, this capability allows for the dynamic adjustment of the phase-shift matrix for each individual slot.

Denoting
\begin{equation}
    \boldsymbol {\Theta}_k=\mathrm{diag} \left\{e^{j\theta_{11}[k] },\ldots, e^{j\theta_{sn_s}[k] },\ldots,e^{j\theta_{SN_S}[k]}\right\}
\end{equation}
as the phase-shift matrix for time slot $k$, with the time-selective phase shift $\theta_{sn}[k]$, where $s=1,\ldots,S$, $n_s=1,\ldots,N_s$, and $k=1,\ldots,K$. Without loss of generality, the received signal for TDMA user $k$ at slot $k$ is expressed by \begin{equation}
    r_k=\Biggl(\sum_{s=1}^{S} \sum_{n_s=1}^{N_s}  f_{sn_s} e^{j\theta_{sn_s}[k]} g_{k,sn_s}  + d_{k}\Biggr)  s_k + n,
\end{equation}
where $n$ is additive white Gaussian noise (AWGN) with zero mean and variance $\sigma_n^2$, i.e. $n\sim \mathcal{CN}(0,\sigma_n^2)$, $d_k$ stands for the coefficient of the direct channel from user $k$ to the BS, $s_k\in \mathbb{C}$ denote the information symbol from user $k$, satisfying $\mathbb{E}[|s_k|^2]\leqslant P_k$, where $P_k$ denotes the power constraint of user $k$ and $\mathbb{E}[\cdot]$ denotes mathematical expectation. Aligning the phase of every reflected signal with that of the LOS signal is essential for coherent signal combining at the receiver. Therefore, it becomes evident that determining the optimal phase-shift matrix involves a straightforward derivation:
\begin{equation} \label{eqn_IRS_phaseshifts}
  \boldsymbol{\Theta}_k^\star =\mathrm{diag}\left \{e^{j\left(\arg\left(d_k\right)-\arg\left( \mathrm{diag}(\mathbf{f})\mathbf{g}_k \right)\right)} \right\},
\end{equation}
where $\arg(\cdot)$ stands for the phase of a complex scalar or vector, and \begin{equation}
    \mathbf{g}_k=\left[\mathbf{g}_{k1}^T,\mathbf{g}_{k2}^T,\ldots,\mathbf{g}_{kS}^T \right]^T,
\end{equation}
\begin{equation}
    \mathbf{f}=\left[\mathbf{f}_{1}^T,\mathbf{f}_{2}^T,\ldots,\mathbf{f}_{S}^T \right]^T.
\end{equation}
It results in a per-user rate of
\begin{equation}
    R_k=\frac{1}{K}\log \left(1+\left| \sum_{s=1}^{S} \sum_{n_s=1}^{N_s}  \left |f_{sn_s}  g_{k,sn_s} \right| +|d_k|\right|^2 \frac{P_k  }{\sigma_n^2}\right).
\end{equation}
Thereby, the sum rate of the TDMA-RIS system can be computed by
\begin{equation} \label{IRS_EQN_TDMA_SE}
    C_{tdma}=\sum_{k=1}^K\frac{1}{K}\log\left(1+\frac{P_k \Bigl| \sum_{s=1}^{S}\sum_{n_s=1}^{N_s}  \left |f_{sn_s}  g_{k,sn_s} \right| +|d_k| \Bigr|^2 }{\sigma_n^2} \right),
\end{equation}
where the factor $1/K$ is due to the orthogonal partitioning of the time resource.

From the information-theoretic perspective, TDMA is inefficient because each user utilizes only a fraction of the resources. Hence,  non-orthogonal transmission, see \cite{Ref_tse2005fundamentals, Ref_jiang20226G, Ref_yuan2016nonorthogonal, Ref_ding2017survey}, are applied.
In this case, the users simultaneously send their signals toward the BS over the same time-frequency resource. The BS observes the received signal of
\begin{equation} \label{eqn_systemModel}
    r=\sum_{k=1}^{K} \Biggl(\sum_{s=1}^{S} \sum_{n_s=1}^{N_s}  f_{sn_s} e^{j\theta_{sn_s}} g_{k,sn_s}  + d_{k} \Biggr) s_k + n.
\end{equation}
Define an overall phase-shift matrix  as
\begin{equation}
    \boldsymbol{\Theta}=\mathrm{diag}\biggl\{\boldsymbol{\Theta}_1,\ldots,\boldsymbol{\Theta}_S\biggr\},
\end{equation}
where
\begin{equation}
    \boldsymbol{\Theta}_s=\mathrm{diag}\{e^{j\theta_{s1}},\ldots,e^{j\theta_{sN_s}}\},
\end{equation}
\eqref{eqn_systemModel} can be rewritten in matrix form as
\begin{equation} \label{EQN_IRS_RxSignal_Matrix}
    r=\sum_{k=1}^{K} \Biggl(\mathbf{f}^T \boldsymbol{\Theta} \mathbf{g}_{k}  + d_{k} \Biggr) s_k + n,
\end{equation}

Our goal is to maximize the \textit{sum capacity}, i.e.,
\begin{equation} \label{EQN_SumRate2}
    C =\log \left(  1 +\frac{ \sum_{k=1}^K \left|\mathbf{f}^T \boldsymbol{\Theta} \mathbf{g}_k +d_{k}\right|^2P_k}{\sigma_n^2}  \right).
\end{equation}
Let $\mathbf{G}=\left[\mathbf{g}_{1},\mathbf{g}_{2}, \ldots, \mathbf{g}_{K}\right]$ and $\mathbf{d}=[d_1,d_2,\ldots,d_K]^T$. It is not difficult to verify that $\sum_{k=1}^K \left|\mathbf{f}^T \boldsymbol{\Theta} \mathbf{g}_k +d_{k}\right|^2=\left\|\mathbf{f}^T \boldsymbol{\Theta} \mathbf{G} +\mathbf{d}^T\right\|^2$.  Assume all users have the same power constraint, i.e., $P_k=P_u$, $\forall k$, \eqref{EQN_SumRate2} is equivalent to
\begin{equation} \label{eqn_IRS_sumRate}
    C=   \log  \left(  1+\frac{  \left\|\mathbf{f}^T \boldsymbol{\Theta} \mathbf{G} +\mathbf{d}^T \right\|^2P_u}{\sigma_n^2}  \right),
\end{equation}
which is a function of $\boldsymbol{\Theta}$, resulting in the optimization formula
\begin{equation}
\begin{aligned} \label{EQN:sumRate_Optimization}
\max_{\boldsymbol{\Theta}} \quad & \left\|\mathbf{f}^T \boldsymbol{\Theta} \mathbf{G} +\mathbf{d}^T\right\|^2\\
\textrm{s.t.} \quad & \theta_{sn_s}\in [0,2\pi), \: \forall s, n_s.
\end{aligned}
\end{equation}

If the system knows CSI well, a set of \textit{optimized} phase shifts can be obtained by solving \eqref{EQN:sumRate_Optimization} directly. As \cite{Ref_wu2019intelligent, Ref_yan2020passive, Ref_wu2020joint}, we can apply the SDR approach to jointly optimize RIS reflection, and a numerical algorithm named CVX \cite{cvx} can figure out the globally optimal solution. Unfortunately, JRO suffers from high computational complexity, blocking its use in practical systems.

To simplify the optimization, we can opportunistically select a user as the unique transmitter.
The \textit{single-user bound} \cite{Ref_tse2005fundamentals} for a typical user $k$, which means the capacity of a point-to-point link with the other users absent from the system,  is given by
\begin{equation} \label{EQN_IRS_SingleUserBound}
  C_k = \log  \left(  1+\frac{\left|\mathbf{f}^T \boldsymbol{\Theta} \mathbf{g}_k +d_{k}\right|^2P_u}{\sigma_n^2}  \right),\: \forall\: k.
\end{equation}
The optimal reflection to maximize $C_k$ is $\mathbf{\Theta}_{k}=\mathrm{diag}\left \{e^{j\left( \arg\left( d_k\right)  -\arg\left( \mathrm{diag}(\mathbf{f})\mathbf{g}_k\right)\right)}\right\}$, achieving coherent combining at the receiver. The resultant capacity equals to $C_k =\log\left(1+\gamma_k\frac{P_k}{\sigma_n^2} \right)$ with
the maximized end-to-end (E2E) channel power gain for user $k$
\begin{equation} \label{EQN_RIS_GammaK}
   \gamma_k =   \left| \sum_{s=1}^{S} \sum_{n_s=1}^{N_s}  |f_{sn_s} || g_{k,sn_s}|  + |d_{k}| \right|^2.
\end{equation}
The philosophy of OUS is to determine the best user as $k^\star=\arg \max_{k\in\{1,\ldots,K\}} \left(\gamma_k\right)$.
Other users turn off when $k^\star$ is transmitting its signal, while the RIS phase shifts are tuned to $\mathbf{\Theta}_{k^\star}$.
The effective E2E power gain of OUS is equal to \begin{equation}  \label{eqn_E2EchannelGain}
  \gamma_{ous} = \max_{k\in\{1,\ldots,K\}} \biggl(\gamma_k\biggr)=\max_{k\in\{1,\ldots,K\}} \left (  \left| \sum_{s=1}^{S} \sum_{n_s=1}^{N_s}  |f_{sn_s} || g_{k,sn_s}|  + |d_{k}| \right|^2 \right).
\end{equation}

\begin{table}[!t]
\renewcommand{\arraystretch}{1.3}
\scriptsize
\caption{Differentiation between the benchmark and proposed schemes.}
\label{table_benck}
\centering
\begin{tabular}{l|c|c|c|c|c}
\cline{2-6}
 &\multicolumn{2}{c|}{\textbf{ Signal Transmission }}& \multicolumn{3}{c}{\textbf{RIS Optimization}}\\ \cline{2-6}
&  Sin. User & Mul.-User& No CSI& Sin. User & Mul.-User \\ \hline
TDMA &$\checkmark$&&&$\checkmark$& \\ \hline
OppBF &$\checkmark$&&$\checkmark$&& \\ \hline
OUS &$\checkmark$&&&$\checkmark$& \\ \hline
JRO &&$\checkmark$&&&$\checkmark$ \\ \hline \hline
OR-NOMA&&$\checkmark$&&$\checkmark$& \\ \hline
OR-NOMA-RP&&$\checkmark$&$\checkmark$&& \\ \hline
\multicolumn{6}{l}{\footnotesize Note: \textit{Sin. User} means a single user selected for transmission or reflection;}\\
\multicolumn{6}{l}{\footnotesize  \textit{Mul.-User} denotes joint transmission or reflection of multiple users.} \\
\end{tabular}
\end{table}


Despite simple optimization, OUS does not fully utilize the resource and suffers from performance degradation.   In \cite{Ref_jiang2023opportunistic}, opportunistic reflection with all-user transmission has been proposed. Although it achieves a good performance-complexity trade-off, it is vulnerable to hardware and channel impairments, such as channel aging and phase noise \cite{Ref_jiang2023performance}. Besides, all-user non-orthogonal transmission is hard to implement since the capability of the SIC at the receiver is limited by the number of superimposed users.
Responding to this, we propose a simple but efficient scheme, which is more practical than opportunistic multi-user reflection in \cite{Ref_jiang2023opportunistic}. The main idea is to opportunistically select a user as the anchor for optimizing the RIS reflection but only two users are allowed to simultaneously transmit, rather than all users.  The best user is determined as $k^\star=\arg \max_{k\in\{1,\ldots,K\}} \left(\gamma_k\right)$.  The RIS phase shifts are adjusted to $\mathbf{\Theta}_{OR-NOMA}= e^{j\left( \arg\left( d_{k^\star}\right)  -\arg\left( \mathrm{diag}(\mathbf{f})\mathbf{g}_{k^\star}\right)\right)}$. Unlike OUS/OppBF, which assigns all resources to the unique transmitter, the second user denoted by $k'$ is selected to superpose its signal on that of the primary user, where the benefits of non-orthogonal transmission, i.e., high system capacity and improved user fairness, are obtained. In this case, the BS observes
\begin{align} \nonumber
    r_{OR-NOMA}
    &=\underbrace{\Biggl(\mathbf{f}^T \boldsymbol{\Theta}_{OR-NOMA} \mathbf{g}_{k^\star}  + d_{k^\star} \Biggr) s_{k^\star}}_{\text{Optimal\:Reflection}} \\
    &+\underbrace{ \Biggl(\mathbf{f}^T \boldsymbol{\Theta}_{OR-NOMA} \mathbf{g}_{k'}  + d_{k'} \Biggr) s_{k'}}_{\text{Non-Orthogonal\:Transmission}} + n.
\end{align}

The receiver first detects $s_{k^\star}$ by treating non-orthogonal transmitted signals as noise.
Applying SIC \cite{Ref_tse2005fundamentals, Ref_jiang20226G, Ref_yuan2016nonorthogonal, Ref_ding2017survey}, the receiver cancels the corresponding interference $(\mathbf{f}^T \boldsymbol{\Theta}_{OR-NOMA} \mathbf{g}_{k^\star}  + d_{k^\star}) s_{k^\star}$ from $r_{OR-NOMA}$ and then detects the symbol from the second user.  It is not hard to derive that the achievable sum rate equals $C_{OR-NOMA} =
    \log \left(  1 +\gamma_{OR-NOMA} \frac{  P_u}{\sigma_n^2}  \right)$
with
\begin{align} \label{eqn_OR-NOMA_gamma} \nonumber
    \gamma_{OR-NOMA}&=\left| \sum_{s=1}^{S} \sum_{n_s=1}^{N_s}  |f_{sn_s} || g_{k^\star,sn_s}|  + |d_{k^\star}| \right|^2\\ &+ \left|\mathbf{f}^T \boldsymbol{\Theta}_{OR-NOMA} \mathbf{g}_{k'} +d_{k'}\right|^2.
\end{align}
The proposed scheme is also depicted in Algorithm 1.  As compared in Table \ref{table_benck}, it differs from either JRO where all users participate in signal transmission and RIS optimization, or OUS/OppBF, which has a single transmitter.

Under fast fading, a quick change of channel imposes a high overhead on CSI acquisition, especially for RIS with massive elements. Inspired by OppBF \cite{Ref_nadeem2021opportunistic, Ref_dimitropoulou2022opportunistic},  we further propose a simplified variant named OR-NOMA with random phase (OR-NOMA-RP) to avoid the high complexity of RIS channel estimation. In the absence of CSI, the RIS elements randomly set their phase shifts.  Two non-orthogonally combined users simultaneously transmit their signals and the receiver applies SIC to deal with inter-user interference, as described by Algorithm \ref{alg:OR-NOMA-RP}.
\SetKwComment{Comment}{/* }{ */}
\RestyleAlgo{ruled}
\begin{algorithm}
\caption{OR-NOMA} \label{alg:IRS002}
\ForEach{Coherence Interval}{
Estimate $\mathbf{f}$, $\mathbf{d}$, and $\mathbf{g}_k$, $\forall k$\;
$k^\star \gets \arg \max_{k\in\{1,\ldots,K\}} \bigl( \gamma_k \bigr)$\;
Tune RISs to $\mathbf{\Theta}_{OR-NOMA}=e^{j\left( \arg\left( d_k^\star\right)  -\arg\left( \mathrm{diag}(\mathbf{f})\mathbf{g}_k^\star\right)\right)}$\;
Randomly or deliberately select another user\;
Two NOMA users simultaneously transmit\;
The receiver conducts SIC\;
}
\end{algorithm}

\SetKwComment{Comment}{/* }{ */}
\RestyleAlgo{ruled}
\begin{algorithm}
\caption{OR-NOMA with Random Phase Shifts} \label{alg:OR-NOMA-RP}
\ForEach{Transmission Block}{
RISs randomly set phase shifts\;
Randomly or deliberately select another user\;
Two NOMA users simultaneously transmit\;
The receiver conducts SIC\;}
\end{algorithm}

\section{Numerical results}

Monte-Carlo simulations are employed to comparatively assess the effectiveness of different strategies. This section first elaborates on the simulation parameters and then provides some representative numerical results.  To establish a representative scenario, we adopt the subsequent simulation configuration: a base station positioned at the center of a circular cell having a radius of \SI{300}{\meter}. Within this cell, $K=4$ users are distributed randomly. Surrounding the BS, four surfaces, each equipped with $N_s=100$ elements, are uniformly deployed on a concentric circle situated at a distance of \SI{60}{\meter}. The minimum power allocated to user equipment is designated as $P_u=0.1\mathrm{W}$ and is gradually increased by a factor ranging from \SIrange{0}{30}{\decibel}. The signal bandwidth is set at $10\mathrm{MHz}$ while the noise power density remains at $-174\mathrm{dBm/Hz}$, accompanied by a noise figure of $9\mathrm{dB}$. The large-scale fading for scenarios without a direct line of sight is determined using the 3GPP Urban Micro (UMi) model, represented as $\Omega = - 22.7 - 26 \log(f_c) - 36.7 \log(d)$, with $d$ being the distance and the carrier frequency set at $f_c=2\mathrm{GHz}$. Taking into account the LOS component, the path loss for the link between the base station  and the RIS is computed as $\mathcal{L}_0/d^{-\alpha}$, where $\mathcal{L}_0=\SI{-30}{\decibel}$ represents the path loss at the reference distance of \SI{1}{\meter}, and the path-loss exponent is denoted as $\alpha=2$. Additionally, a Nakagami-$m$ fading is generated with the parameter $m$ set to $2.5$.

\begin{figure}[!t]
    \centering
    \includegraphics[width=0.49\textwidth]{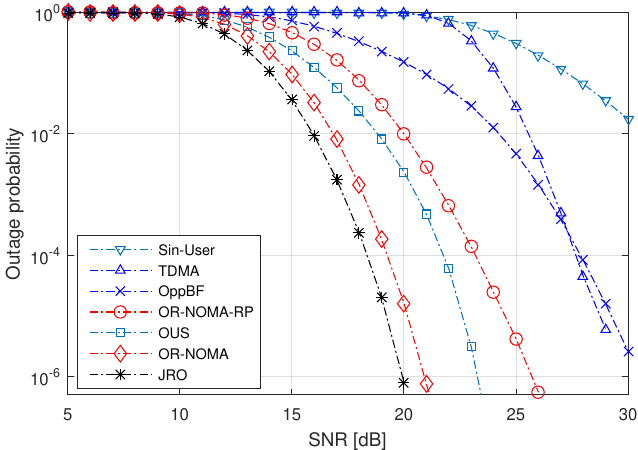}
    \caption{Performance comparison of different schemes. For clear illustration, the target rate is set to $R=\SI{8}{\bps\per\hertz^{}}$.  }
    \label{Fig_outagePerformance}
\end{figure}

\figurename \ref{Fig_outagePerformance} provides a comprehensive comparison among different schemes, including 1) the single-user case, 2) OppBF, 3) TDMA, 4) OUS, 5) OR-NOMA, 6) OR-NOMA with random phase shifting, and 7) JRO.
Benefiting from the multiplexing gain, all the multi-user transmission methods—joint, opportunistic, and round-robin—outperform single-user transmission. With the availability of CSI, the proposed approach achieves performance levels closely approximating those of JRO, significantly superior to TDMA and OUS. In cases without CSI knowledge, the proposed method exhibits an SNR improvement of approximately \SI{5}{\dB} over OppBF. Last but not least,  the computational complexity of the three strategies with the availability of CSI is assessed based on the average computer central processing unit (CPU) runtime per channel realization. The evaluation is conducted on a platform employing an Intel i7-4790 CPU operating at 3.60GHz, accompanied by 32GB of memory. The computations are executed using Matlab parallel computing with 4 workers. As obtained in the simulations, JRO, due to its employment of SDR, demonstrates a complexity of $2.39\times 10^4\si{\milli \second}$ per channel realization that is \textit{four orders of magnitude higher} than that of the opportunistic approaches, amounting to $0.83\si{\milli \second}$ and $1.43\si{\milli \second}$ for OppBF and OR-NOMA, respectively. Given that the channel coherence time typically falls within the range of \SIrange{10}{100}{\milli \second}, the computational demands of JRO render it infeasible for practical applications, while the efficiency of OR-NOMA remains moderate.

\section{Conclusions}
In this paper, we proposed a simple but efficient transmission approach, coined opportunistic reflection-based non-orthogonal multiple access or OR-NOMA, for multi-user multi-RIS-aided systems. The key idea is opportunistically selecting the best user that can reap the maximal gain from RIS reflection, and then the second user is selected to superpose its signal on that of the primary user. With the availability of CSI, OR-NOMA achieves a near-optimal performance close to JRO but its complexity is as low as OUS, which substantially simplifies RIS optimization by several orders of magnitude.   When the CSI is hard to acquire, a simplified variant named OR-NOMA-RP exploiting random phase shifts is proposed to avoid the high overhead of RIS channel estimation.  OR-NOMA-RP remarkably outperforms the benchmark OppBF scheme.

\bibliographystyle{IEEEtran}
\bibliography{IEEEabrv,Ref_CommL}

\end{document}